\newif\ifREFEREE
 \REFEREEtrue
 \REFEREEfalse

\ifREFEREE
        \documentstyle[epsf,referee]{mn}
\else
        \documentstyle[epsf]{mn}
\fi

\newif\ifAMStwofonts
\AMStwofontsfalse

\def\fracj#1#2{{\textstyle{#1\over#2}}}
\def\pdot{\,\dot+\,}

\def\hip{{\it Hipparcos\/}}
\def\tyc{{\it Tycho\/}}

\def\ssy{\scriptstyle}

\def\bea{\begin{array}}
\def\eea{\end{array}}
\def\beq{\begin{equation}}
\def\eeq{\end{equation}}
\def\ben{\begin{eqnarray}}
\def\een{\end{eqnarray}}

\renewcommand{\d}{\rmn{d}}
\def\p{\upartial}
\def\pr{\prime}
\renewcommand{\b}[1]{\bmath{#1}}
\renewcommand{\d}{\rmn{d}}
\newcommand{\bA} {{\bf A}}
\newcommand{\bB} {{\bf B}}
\newcommand{\bvp}{\b{v}^\pr}
\newcommand{\bv} {\b{v}}
\newcommand{\br} {\b{r}}
\newcommand{\bp} {\b{p}}
\newcommand{\bu} {\b{u}}
\newcommand{\bs} {\b{s}}
\newcommand{\bpp}{\b{p}^\pr}

\newcommand{\pp}{p^\pr}
\newcommand{\bsq}{\bsigma^2}
\newcommand{\ave}[1]{\left\langle{#1}\right\rangle}
\newcommand{\avet}[1]{\langle{#1}\rangle}
\newcommand{\aves}[1]{{\ssy\langle}{#1}{\ssy\rangle}}

\def\twothirds{\hbox{$\,{}^2\!/_3$}}
\def\spose#1{\hbox to 0pt{#1\hss}}
\def\lta{\mathrel{\spose{\lower 3pt\hbox{$\mathchar"218$}}
     \raise 2.0pt\hbox{$\mathchar"13C$}}}
\def\gta{\mathrel{\spose{\lower 3pt\hbox{$\mathchar"218$}}
     \raise 2.0pt\hbox{$\mathchar"13E$}}}

\def\kms{\mbox{$\,{\rm km}\,{\rm s}^{-1}$}}
\def\kpc{\mbox{$\,{\rm kpc}$}}

\def\etal{\mbox{\it et~al.}}

\def\refer#1#2#3{#1, #2, #3}

\def\aaa#1#2{\refer{A\&A}{#1}{#2}}

\def\apj#1#2{\refer{ApJ}{#1}{#2}}

\def\mn#1#2{\refer{MNRAS}{#1}{#2}}

\ifoldfss
  \newcommand{\rmn}[1] {{\rm #1}}

  \ifCUPmtlplainloaded \else
    \NewTextAlphabet{textbfit} {cmbxti10} {}
    \NewTextAlphabet{textbfss} {cmssbx10} {}
    \NewMathAlphabet{mathbfit} {cmbxti10} {} 
    \NewMathAlphabet{mathbfss} {cmssbx10} {} 
  \fi
  \ifAMStwofonts
    \ifCUPmtlplainloaded \else
      \NewSymbolFont{upmath} {eurm10}
      \NewSymbolFont{AMSa} {msam10}
      \NewMathSymbol{\upi}     {0}{upmath}{19}
      \NewMathSymbol{\umu}     {0}{upmath}{16}
      \NewMathSymbol{\upartial}{0}{upmath}{40}
      \NewMathSymbol{\leqslant}{3}{AMSa}{36}
      \NewMathSymbol{\geqslant}{3}{AMSa}{3E}

       \let\le=\leqslant
       
    \fi
  \fi
\fi 

\ifnfssone
  \newmathalphabet{\mathit}
  \addtoversion{normal}{\mathit}{cmr}{m}{it}
  \addtoversion{bold}{\mathit}{cmr}{bx}{it}
  \newcommand{\rmn}[1] {\mathrm{#1}}

  \def\textbfit{\protect\txtbfit}
  
  \long\def\txtbfit#1{{\fontfamily{cmr}\fontseries{bx}\fontshape{it}%
    \selectfont #1}}

  \newmathalphabet{\mathbfit} 
  \addtoversion{normal}{\mathbfit}{cmr}{bx}{it}
  \addtoversion{bold}{\mathbfit}{cmr}{bx}{it}
  \newmathalphabet{\mathbfss} 
  \addtoversion{normal}{\mathbfss}{cmss}{bx}{n}
  \addtoversion{bold}{\mathbfss}{cmss}{bx}{n}
  \ifAMStwofonts
    \ifCUPmtlplainloaded \else
      \UseAMStwoboldmath
      \makeatletter
      \new@mathgroup\upmath@group
      \define@mathgroup\mv@normal\upmath@group{eur}{m}{n}
      \define@mathgroup\mv@bold\upmath@group{eur}{b}{n}
      \edef\UPM{\hexnumber\upmath@group}
      \new@mathgroup\amsa@group
      \define@mathgroup\mv@normal\amsa@group{msa}{m}{n}
      \define@mathgroup\mv@bold\amsa@group{msa}{m}{n}
      \edef\AMSa{\hexnumber\amsa@group}
      \makeatother
      \mathchardef\upi="0\UPM19
      \mathchardef\umu="0\UPM16
      \mathchardef\upartial="0\UPM40
      \mathchardef\leqslant="3\AMSa36
      \mathchardef\geqslant="3\AMSa3E

       \let\le=\leqslant

    \fi
  \fi
\fi 

\ifnfsstwo
  \newcommand{\rmn}[1] {\mathrm{#1}}

  \def\textbfit{\protect\txtbfit}
  
  \long\def\txtbfit#1{{\fontfamily{cmr}\fontseries{bx}\fontshape{it}%
    \selectfont #1}}

  \DeclareMathAlphabet{\mathbfit}{OT1}{cmr}{bx}{it}
  \SetMathAlphabet\mathbfit{bold}{OT1}{cmr}{bx}{it}
  \DeclareMathAlphabet{\mathbfss}{OT1}{cmss}{bx}{n}
  \SetMathAlphabet\mathbfss{bold}{OT1}{cmss}{bx}{n}
  \ifAMStwofonts
    \ifCUPmtlplainloaded \else
      \DeclareSymbolFont{UPM}{U}{eur}{m}{n}
      \SetSymbolFont{UPM}{bold}{U}{eur}{b}{n}
      \DeclareSymbolFont{AMSa}{U}{msa}{m}{n}
      \DeclareMathSymbol{\upi}{0}{UPM}{"19}
      \DeclareMathSymbol{\umu}{0}{UPM}{"16}
      \DeclareMathSymbol{\upartial}{0}{UPM}{"40}
      \DeclareMathSymbol{\leqslant}{3}{AMSa}{"36}
      \DeclareMathSymbol{\geqslant}{3}{AMSa}{"3E}

       \let\le=\leqslant

    \fi
  \fi
\fi 

\ifCUPmtlplainloaded \else
  \ifAMStwofonts \else 
    \def\upi{\pi}
    \def\umu{\mu}
    \def\upartial{\partial}
  \fi
\fi

\begin{document}

 \title[Local stellar kinematics]
       {Local stellar kinematics from \textbfit{Hipparcos} data}

 \author[W.\ Dehnen and J.\ Binney]
        {Walter Dehnen and James J.~Binney  \\
 Theoretical Physics, 1 Keble Road, Oxford OX1 3NP}

 
\pubyear{1997}

\maketitle

\begin{abstract}
 From the parallaxes and proper motions of a kinematically unbiased
subsample of the \hip\ catalogue we have redetermined as a function of colour
the kinematics of main-sequence stars.

Whereas the radial and vertical components of the mean heliocentric velocity
of stars show no trend with colour, the component in the direction of
galactic rotation nicely follows the asymmetric drift relation, except for
stars bluer than $B$--$V{\,=\,}0.1\,$mag. Extrapolating to zero dispersion
yields for the velocity of the Sun w.r.t.\ the local standard of rest (LSR)
in $\kms$: $U_0{\,=\,}10.00\pm0.36$ (radially inwards), $V_0{\,=\,}
5.23\pm0.62$ (in direction of galactic rotation), and
$W_0{\,=\,}7.17\pm0.38$ (vertically upwards).

Parenago's discontinuity is beautifully visible in a plot of velocity
dispersion against colour: the dispersion, which is essentially constant for
late spectral types, decreases towards early spectral types blueward of
$B-V\approx0.62\,$mag.

We determine the velocity-dispersion tensor $\bsq$ as function of colour.
The mixed moments involving vertical motion are zero within the errors,
while $\sigma^2_{xy}$ is non-zero at about $(10\kms)^2$ independent of
colour. The resulting vertex deviations are about 20 degrees for early-type
stars and $10\pm4$ degrees for old-disc stars. The persistence of the vertex
deviation to late-type stars implies that the Galactic potential is
significantly non-axisymmetric at the solar radius. If spiral arms are
responsible for the non-axisymmetry, they cannot be tightly wound.

Except for stars bluer than $B$--$V{\,=\,}0.1\,$mag the ratios of the 
principal velocity dispersions are given by $\sigma_1{\,:\,}\sigma_2
{\,:\,}\sigma_3{\,\approx\,}2.2{:}1.4{:}1$, while the absolute values 
increase with colour from $\sigma_1{\,\approx\,}20\kms$ at $B$--$V{\,=\,}
0.2\,$mag to $\sigma_1{\,\approx\,}38\kms$ at Parenago's discontinuity 
and beyond. These ratios imply significant heating of the disc by spiral
structure and that $R_0/R_\d\simeq3$ to $3.5$, where $R_\d$ is the scale 
length of the disc.

\end{abstract}


\section{INTRODUCTION}
The kinematics of stars near the Sun has long been known to provide crucial
information regarding both the structure and the evolution of the Milky Way.
Karl Schwarzschild (1908) already interpreted the distribution of random
velocities as forming a triaxial `velocity ellipsoid', which Oort, B.~Lindblad
and Str\"omberg were able to relate to the large-scale structure of the disc.
In the early 1950s Parenago (1950), Nancy Roman (1950, 1952) and others pointed
out that stellar kinematics varies systematically with stellar type, in the
sense that groups of stars that are on average younger have smaller velocity
dispersions and larger mean Galactic rotation velocities than older stellar
groups. Spitzer \& Schwarzschild (1953), Barbanis \& Woltjer (1967), and 
Wielen (1977) explained these correlations in terms of the diffusion of stars
through phase space as the Galactic disc ages -- for recent studies of these 
processes see Binney \& Lacey (1988) and Jenkins (1992).

The \hip\ Catalogue \cite{HipTyc} provides an important opportunity to 
re-examine the fundamental data of solar-neighbourhood kinematics by providing
the first all-sky catalogue of {\em absolute\/} parallaxes and proper motions.
From the \hip\ Catalogue we can, moreover, extract samples that are
completely free of the kinematic biases that have plagued similar studies in 
the past. Binney et al.\ (1997) obtained some preliminary results from a
sample of 5610 stars around the south celestial pole. Here we extend these
results in three ways:

\newcounter{bean}
\begin{list}{\arabic{bean}.\ }{\usecounter{bean}}%
	\setlength{\rightmargin}{\leftmargin}

\item We have joined  the relatively deep sample of Binney et al.\
(1997) to a sample that covers the whole sky but has a significantly brighter
limiting magnitude. The combined sample gives significantly better
statistics for earlier spectral types.

\item Since we now have some all-sky coverage we are able to impose a
stricter limit on parallax errors: we use only stars with
$\sigma_\pi/\pi\le0.1$, whereas Binney et al.\ accepted stars with
$\sigma_\pi/\pi\le0.2$. 

\item Whereas Binney et al.\ only reported the solar motion and a single
estimate of random velocity for each spectral type, we derive the lengths of
the axes of the velocity ellipsoid and analyze its orientation.

\end{list}

Section 2 describes the new sample. Section 3 explains how we have  analyzed
the sample. Section 4 gives the results and Section 5 discusses them.

\ifREFEREE
\else
\begin{figure}
        \epsfxsize=21pc \epsfbox[18 170 500 720]{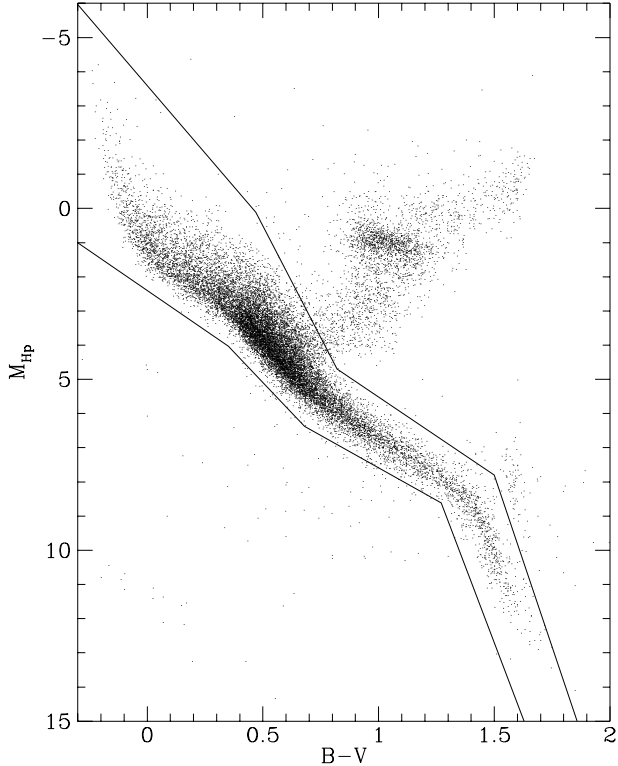}
 \caption[]{Hertzsprung-Russell diagram ($M_{Hp}$ is the absolute magnitude in
	    \hip's own passband) of the 18\,860 single \hip\ stars with
	    relative parallax errors less than 10 per cent. The lines are used
	    to select the main sequence and have 16\,054 stars between them.}
 \label{fig-HR-acc}
\end{figure}
\fi

\section{THE SAMPLE}

Our sample has to meet several criteria if it is accurately to determine
local stellar kinematics as a function of stellar age. First, it must be
kinematically unbiased, i.e., faithfully represent at each point in the
Hertzsprung-Russell diagram the kinematics of all nearby stars. Second, it
should be confined to main-sequence stars, since only for these is there a
one-to-one relation between age and either colour or absolute luminosity.
Third, it should be based on accurate astrometry, and, fourth, it should
contain no multiple stars because their kinematics contains additional
motions. 

We ensured satisfaction of the last two criteria by taking only single stars
with relative parallax errors smaller than 10 per cent. This selection
criterion automatically excludes remote stars, and leaves just 18\,860 of the
118\,218 stars in the \hip\ catalog. The Hertzsprung-Russell diagram of the
selected stars is shown in Figure~\ref{fig-HR-acc}, which also shows the
division lines that we used to select main-sequence stars; there are 16\,054
stars between these lines.  However, given the very heterogeneous nature of
the \hip\ catalog, they  almost certainly do not comprise a kinematically
unbiased sample. The only reliable way to extract such a sample is to
isolate a magnitude-limited subsample. 

According to its Volume I, p.~4, the \hip\ Catalogue comprises some 60\,000
objects complete to about $7.3$ to $9\,$mag depending on galactic latitude,
$b$, and stellar type. Actually, however, 59 stars brighter than $V_T=7.0$
(the subscript $T$ denotes the passbands used by \tyc, the starmapper of
\hip) are absent from the Catalogue (see the table on p.~142 of Volume I).
We used the \tyc\ Catalogue \cite{HipTyc}, which is essentially complete
to 11\,mag, to construct an almost complete subsample of \hip\ stars.
For each of $16\times16 \times10$ uniformly spaced bins in $\sin b$,
galactic longitude, $\ell$, and $B_T-V_T$, we took all \hip\ stars brighter
(in $V$) than the second brightest star per bin that is contained in the
\tyc\ but not the \hip\ Catalogue. This gives a subsample with 95 per cent
completeness and 47\,558 stars (compared to 60\,000 claimed), of which
10\,706 have $\sigma_\pi/\pi\le0.1$ and lie between the lines on
Figure~\ref{fig-HR-acc}. 

\ifREFEREE
\else
\begin{figure}
        \epsfxsize=21pc \epsfbox[18 200 500 720]{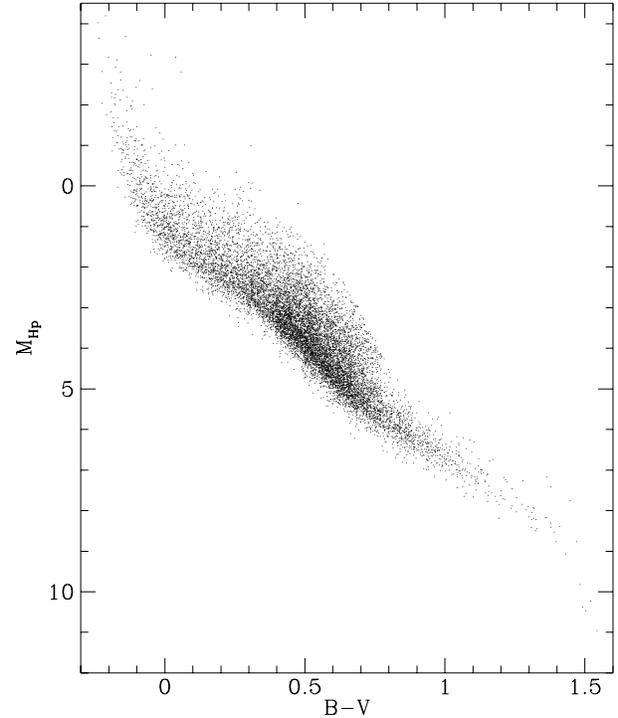}
 \caption[]{Hertzsprung-Russell diagram for the stars in our kinematical unbiased
	    sample of 11\,865 single main-sequence stars.}
 \label{fig-HR-sam}
\end{figure}
\fi

Another kinematically unbiased subsample is given by \hip\ Proposal 018: all
stars spectrally classified in the Michigan Catalogue by 1982
\cite{hc75,houk78,houk82} that, judged from their spectral classification,
should be within 80 pc from the Sun -- this `SCP' sample is the one analyzed
by Binney et al.\ (1997). It contains 6845 stars south of $\delta=-28$
degree (covering 26.5 per cent of the sky), of which 3172 are single and have
$\sigma_\pi/\pi\le0.1$ and lie between the lines on Figure~\ref{fig-HR-acc}.

Whereas most of the stars in the all-sky sample are of early type and
intrinsically luminous, the SCP sample contains many late-type dwarfs.  The
union of the samples is kinematically unbiased and has 11\,865 entries. It
is this combined sample that we analyze below. Figure~\ref{fig-HR-sam} shows
its Hertzsprung-Russell diagram.

\section[]{ANALYSIS TECHNIQUE}
 \hip\ provides us with parallax measurements of unprecedented accuracy, and
proper motions whose accuracy is only comparable to that of ground-based
astrometry on account of short duration ($3.3\,$yr) of the \hip\ mission.
Unfortunately, the \hip\ astrometry mission was not complemented by a
programme to measure the radial velocities of the same stars, so we do not
know the space velocities of most sample stars. As Binney et al.\ (1997)
have demonstrated, the stars with known radial velocities form a
kinematically biased subsample. Hence we have to discard the known radial
velocities and work with the \hip\ data alone. 

\subsection{The projection equation} \label{sec-proj}
 The velocity of a star relative to the Sun can be divided into three
components: (i) the peculiar velocity of the star, (ii) that of the Sun, and
(iii) the contribution from Galactic rotation. Here we are dealing with the
first two parts, and correct for Galactic rotation before further analysis.
Given observed value $\mu_{\ell}^{(\rmn{obs})}$ and $\mu_{b}^{(\rmn{obs})}$,
the corrected values are 
 \beq \bea{r@{\,=\,}l} 
 \mu_\ell & \mu_{\ell}^{(\rmn{obs})} - A \cos(2\ell)  - B \\ 
 \mu_b & \mu_{b}^{(\rmn{obs})} + A \sin(2\ell)\cos b \sin b.
\eea \eeq 
 We used the values of $A$ and $B$  derived by Feast \& Whitelock
(1997) from \hip\ Cepheids: $A=14.82 \kms\kpc^{-1}$ and
$B=-12.37\kms\kpc^{-1}$. 

 Let us introduce a Cartesian coordinate system such that $\hat\b{x}$ points
towards the Galactic centre, $\hat\b{y}$ in the direction of Galactic
rotation, and $\hat\b{z}$ towards the north Galactic pole. We consider a
star with parallax $\pi$ and heliocentric velocity $\bv$, and define the
star's proper-motion velocity $\bp$ to be
 \beq \bp \equiv{1\over\pi} \left[\bea{r}
	 -\sin\ell\,\cos b\,\mu_\ell - \cos\ell\,\sin b\,\mu_b \\
	  \cos\ell\,\cos b\,\mu_\ell - \sin\ell\,\sin b\,\mu_b \\
					         \cos b\,\mu_b
		   \eea \right].
 \eeq
 Then $\bp$ and $\bv$ are related by
 \beq \label{v-proj}
	\bp = \bA \cdot \bv,
 \eeq
where the matrix $\bA$ is defined by
 \beq \label{proj-matrix}
	\bA\equiv\b{I}-\hat\br\otimes\hat\br,
 \eeq
 with $\hat\br$ the unit vector to the star. $\bA$ projects
velocities onto the celestial sphere. It is symmetric and, as every
projection operator, obeys $\bA^2=\bA$ and is singular. Hence we cannot
invert (\ref{v-proj}); we needed the radial velocity to recover $\bv$.

\subsection{The mean motion}
 The solar motion relative to any given group of stars is simply minus the
mean motion of that group with respect to the Sun, $\bv_\odot =- \ave{\bv}$. 
For a kinematically unbiased sample of nearby stars we can safely assume that
the positions on the sky $\hat\br$ are uncorrelated with the velocities
$\bv$.  With this assumption, taking the sample mean of
Equation~(\ref{v-proj}) yields $\ave{\bp} = \ave{\bA}\cdot\ave{\bv}$, which
can be inverted to give
 \beq \label{vsun-from-pm}
	-\bv_\odot = \ave{\bv} = \ave{\bA}^{-1} \cdot \ave{\bp}.
 \eeq
 It is easy to show that for a group of stars that is isotropically
distributed over a celestial hemisphere $\ave{\bA} = \twothirds \b{I}$, where
the factor $\twothirds$ arises from the fact that we only know two of the
three velocity components for each star. Let us denote the motions relative
to the mean by
 \beq \label{defspprime}\bea{r@{\;\equiv\;}l}
	\bvp & \bv - \ave{\bv} \\
	\bpp & \bp - \bA\cdot\ave{\bv} 
     \eea 
 \eeq
and consider the quantity 
 \beq \label{Sq-def}
     S^2 \equiv \ave{|\bpp|^2}.
 \eeq
 It can be shown that the choice of $\ave{\bv}$ given by (\ref{vsun-from-pm}) 
minimizes $S^2$, and (with Equations~\ref{v-proj} and \ref{proj-matrix}) that
 \beq \label{S-dispersion}
	S^2 = \ave{|\bvp|^2} - \ave{(\bvp\cdot\hat\br)^2}.
 \eeq
Hence $S^2$ is a measure for the velocity dispersion of the group. 

 The variance of $\ave{\bv}$ can be estimated to be
 \beq \label{v-variance}
	\b{V}\left(\ave{\bv}\right) = N^{-1} \ave{\bA}^{-1} S^2,
 \eeq
 where $N$ is the number of stars in the sample, while an estimate for the
variance of $S^2$ is
 \beq
	V(S^2)  = N^{-1} \left(\ave{|\bpp|^4} - \ave{|\bpp|^2}^2\right).
 \eeq
 These error estimates only account for the Poisson noise due to the finite
number of stars, and measurement uncertainties in $\bp$ have been neglected. 

\ifREFEREE
\else
\begin{figure*}
 \centerline { \epsfxsize=28pc \epsfbox[18 363 600 720]{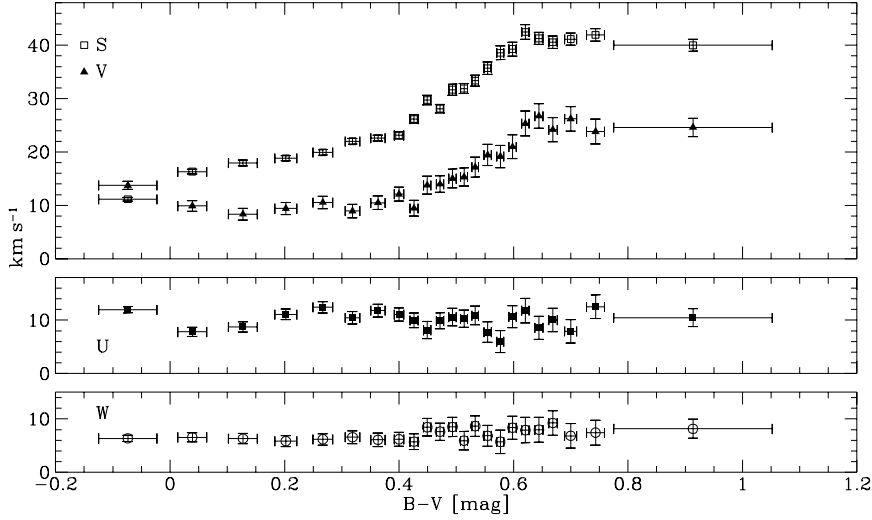} }
 \caption[]{The components $U$, $V$, and $W$ of the solar motion w.r.t.\ stars
	    with different colour $B$-$V$. Also shown is the variation of the
	    dispersion $S$ with colour.}
 \label{fig-mean-motion}
\end{figure*}
\fi

\subsection{The velocity dispersion tensor} \label{sec-tec-sig}
 Similar to the mean velocity, the second-order moments of the space
velocities can be inferred. From Equation~(\ref{v-proj}) we can obtain
 \ben
	\bpp\otimes\bpp &=& (\bA\cdot\bvp) \otimes (\bA\cdot\bvp) \nonumber\\
 \label{vv-proj}
			&=& (\bA\otimes\bA) \cdot (\bvp\otimes\bvp). 
 \een
 The velocity dispersion tensor $\bsq$ is the sample average of $(\bvp\otimes
\bvp)$, so after averaging (\ref{vv-proj}) we can solve for $\bsq$,
exactly as we did for $\ave{\bv}$ above, and the same holds, in
principle, for the velocity moments of higher orders. 

However, if we were to solve the expectation of Equation~(\ref{vv-proj}) for
$\bsq$, we would in general find that the recovered tensor $\bsq$ was not
symmetric, because $\ave{A_{ij}A_{kl}}\neq\ave{A_{kj}A_{il}}$. Since we know
a priori that $\bsq$ {\em is} a symmetric tensor, with six rather than nine
independent elements, it is advantageous to impose this symmetry by solving 
 \beq
	\ave{\pp_i\pp_k} = \fracj12 \, \sum_{jl}\,
	 \ave{A_{ij}A_{kl} + A_{kj}A_{il}} \; \sigma_{jl}.
 \eeq
 For the computer this is more conveniently written as $\ave{\bu} = \bB
\cdot\ave{\bs}$, where $\ave{\bu}$ and $\ave{\bs}$ are vectors containing
the six independent components of $\ave{\pp_i\pp_j}$ and $\sigma_{ij}$,
respectively, whereas $\bB$ is the corresponding $6\times6$ matrix. With
this notation the dispersion tensor is estimated via
 \beq
	\ave{\bs} = \ave{\bB}^{-1}\cdot\ave{\bu},
 \eeq
while an estimate for its variance is given by
 \beq \label{sig-var}
	\b{V}(\ave{\bs}) = N^{-1} \ave{\bB}^{-1} 
		\ave{|\bu - \bB\cdot\ave{\bs}|^2}.
 \eeq
 Somewhat surprisingly, for an isotropic sample the matrix $\ave{\bB}^{-1}$
is not diagonal, in contrast to $\ave{\bA}^{-1}$. It turns out that the
diagonal terms of $\bsq$ are coupled. Quantitatively,
 \beq
 \left(\bea{c} \sigma_x^2 \\ \sigma_y^2 \\ \sigma_z^2 \eea \right)
 = {3\over14} \left(\bea{rrr} 9&-1&-1 \\ -1&9&-1 \\ -1&-1&9 \eea \right) \cdot
 \left(\bea{c}\aves{\pp_x\pp_x}\\ \aves{\pp_y\pp_y}\\ \aves{\pp_z\pp_z} \eea
 \right),
 \eeq
 while all other off-diagonal terms of $\ave{\bB}^{-1}$ average to zero for
an isotropic sample. 

\section[]{RESULTS}
 We will now apply the formul\ae\ from the last section to our sample. The
analysis technique relies heavily on taking averages of the various observed
kinematic quantities. Unfortunately, averages are sensitive to outliers such
as halo stars, and rejection of them is an important issue. Here, we will
use an iterative method: in the solution for the various quantities we
rejected stars that contribute to $S^2$ (Equation~\ref{Sq-def}) by more than
$\kappa$ times the value of $S^2$ obtained in the previous iteration.
Convergence is usually obtained after a few iterations (or never if $\kappa$
is too small).  The results prove to be insensitive to $\kappa$ in the range
of 3 to 4; the results below were obtained with $\kappa=4$.


\ifREFEREE
\else
\begin{figure*}
 \centerline { \epsfxsize=28pc \epsfbox[18 363 600 720]{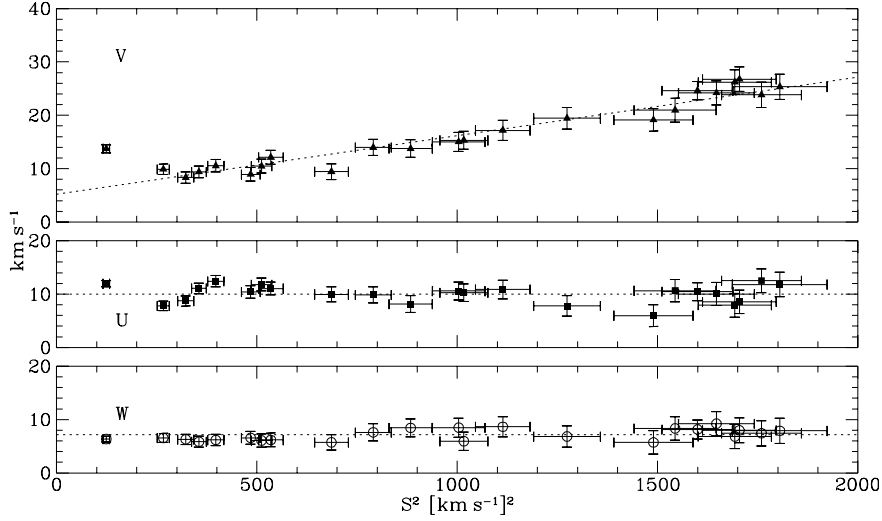} }
 \caption[]{The dependence of $U$, $V$, and $W$ on $S^2$. The dotted lines
	    correspond to the linear relation fitted ($V$) or the mean
	    values ($U$ and $W$) for stars bluer than $B$-$V$=0.}
 \label{fig-solar-motion}
\end{figure*}
\fi

\subsection{The solar motion and Parenago's discontinuity}
 We have binned the stars in $B$-$V$, such that the bins are no
smaller than 0.02\,mag and contain no less than 500 stars.
For each bin, Figure~\ref{fig-mean-motion} plots the solar motion with respect
to the stars in the bin, $\bv_\odot=-\ave{\bv}$, and $S$, which is
a measure for the bin's velocity dispersion, versus the mean colour. As
usual, $U$, $V$, and $W$ denote the components of $\bv_\odot$ in the
$\hat\b{x}$, $\hat\b{y}$, and $\hat\b{z}$ directions as defined in
Section~\ref{sec-proj}. $U$ and $W$ do not vary significantly between bins,
while both $V$ and $S$ increase systematically from early to late spectral
types. The points in $S$ display very beautifully Parenago's (1950) 
discontinuity: around $B$-$V \simeq 0.62\,$mag there is an abrupt change in
gradient from a strongly positive value to about zero. The same discontinuity
is visible, though less clearly, in the data for $V$. Parenago's
discontinuity is thought to arise from the fact that the mean age of stars
decreases as one moves blueward of the discontinuity, while it is independent
of colour redward of the discontinuity: scattering processes cause the random
velocities of stars to increase steadily with age (e.g.\ Jenkins 1992). Hence
velocity dispersion reflects age, and decreases as one moves blueward from
the discontinuity through ever younger stellar groups, while remaining
constant with mean age redward of the discontinuity. The discontinuity itself
should occur at the colour for which the main-sequence lifetime of a star
equals the age of the Galactic disc. However, since stars change colour
during their life on the main sequence, detailed modelling of stellar
populations is necessary, to infer the age of the stellar disc from this
datum (Binney \& Dehnen, in preparation).

\subsection{The Sun's velocity w.r.t.\ the LSR}
 Figure~\ref{fig-solar-motion} is a plot of $U$, $V$ and $W$ versus $S^2$.
For $S\ga15\kms$, this clearly shows the linear
dependence of $V$ on $S^2$ that is predicted by Str\"omberg's asymmetric
drift equation [e.g.\ Equation~(4-34) of Binney \& Tremaine, 1987 (hereafter
BT)]. That is,
$V$ increases systematically with $S^2$ because the larger a stellar group's
velocity dispersion is, the more slowly it rotates about the Galactic centre
and the faster the Sun moves with respect to its lagging frame.

 For very early-type stars with $B$-$V\la0.1\,$mag and/or $S\la15\kms$, the
$V$-component of $\ave{\bv}$ decreases with increasing $S$, colour, and
hence age contradicting the explanation given in the last paragraphs.
However, the stars concerned are very young, and there are several
possibilities for them not to follow the general trend. First, because of
their youth these stars are unlikely to constitute a kinematically
well-mixed sample, rather they move close to the orbit of their parent
cloud; many will belong to a handful of moving groups. Second, Str\"omberg's
asymmetric drift relation predicts a linear relation between $V$ and $S^2$
only if both the shape of the velocity ellipsoid, i.e.\ the ratios of the
eigenvalues of $\bsigma$, and the radial density gradient are independent of
$S$. Young stars probably violate these assumptions, especially the latter
one.

 The velocity $(U_0,V_0,W_0)$ of the Sun with respect to the local standard of
rest (the velocity of the closed orbit in the plane that passes through the
location of the Sun) may be read off from Figure~\ref{fig-solar-motion} by 
extrapolating any trends of $U$, $V$ and $W$ with $S^2$ back to $S=0$. 
Ignoring stars blueward of $B$-$V=0\,$mag we find
 \beq \label{solar-vel}
	\bea{l@{\;=\;}r@{\;\pm\;}rl}
	U_0 & 10.00 & 0.36 & \kms \\
	V_0 &  5.23 & 0.62 & \kms \\
	W_0 &  7.17 & 0.38 & \kms.
 \eea \eeq
 When we use this value of the solar motion to calculate values of
$\avet{v_y}$ relative to the LSR for our stellar groups, Str\"omberg's 
asymmetric drift equation is found to be
 \beq\label{Strom_reln}
\ave{v_y}=-\sigma_{xx}^2/k
\eeq
where $k=80\pm5\kms$ and we have used $S^2\simeq 1.14\sigma_{xx}^2$.

\ifREFEREE
\else
\begin{figure}
        \epsfxsize=21pc \epsfbox[18 150 400 720]{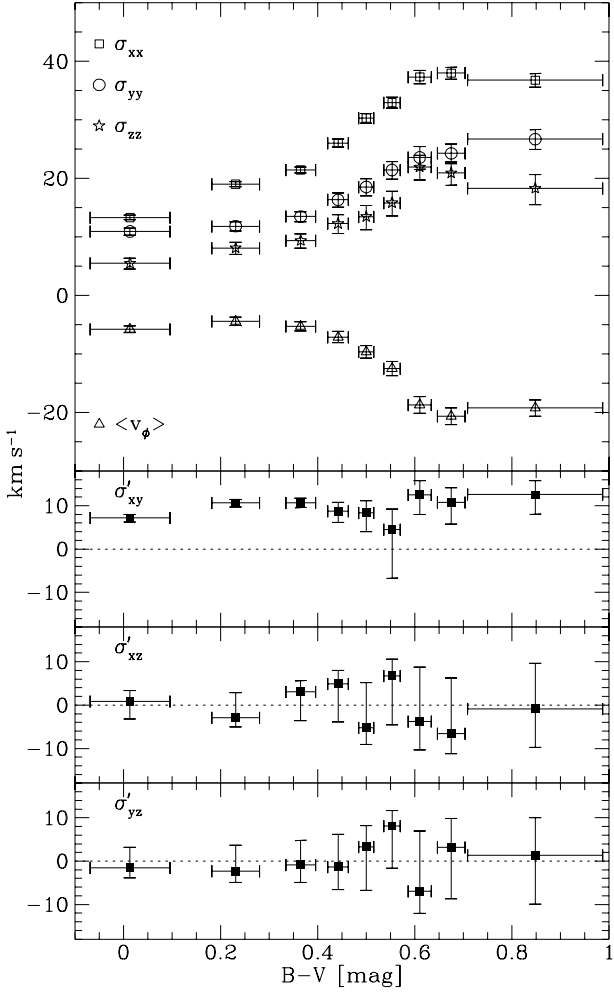}
 \caption[]{Velocity dispersions for stars in different colour bins. The top
	    panel shows the mean rotation velocity (negative values imply
	    lagging w.r.t.\ LSR) and the three main velocity dispersions.
	    In the three bottom panels $\sigma^\pr_{ij}\equiv\rmn{sign}
	    (\sigma^2_{ij})\,|\sigma^2_{ij}|^{1/2}$ is plotted for the mixed
	    components of the tensor $\sigma^2_{ij}$.}
 \label{fig-sigma}
\end{figure}
\fi

\begin{table}
\tabcolsep1mm
\caption[]{Eigenvalues of $\bsq$ for the nine colour bins and all stars beyond
	   Parenago's discontinuity (last row).}\label{tab-sig}
 \begin{tabular}{crr@{\quad}cccc}
   bin  & \multicolumn{2}{c}{$(B$-$V)_\rmn{min,max}$} & $\sigma_1$ & 
	$\sigma_1/\sigma_2$ & $\sigma_1/\sigma_3$ & $\ell_v$ \\
 \hline
   1 &$-0.238$& 0.139 &$ 14.40_{-0.40}^{+0.49} $&$ 1.52_{-0.14}^{+0.16} $&$
	2.62^{+0.91}_{-0.28} $&$ 30.3_{-5.3}^{+4.7}$	\\[0.5ex]
   2 & 0.139 & 0.309 &$ 20.23_{-0.43}^{+0.50} $&$ 2.10_{-0.28}^{+0.13} $&$
	2.50_{-0.10}^{+0.81} $&$ 22.8_{-3.0}^{+2.8}$	\\[0.5ex]
   3 & 0.309 & 0.412 &$ 22.40_{-0.47}^{+0.56} $&$ 1.88_{-0.20}^{+0.13} $&$
	2.39_{-0.14}^{+0.65} $&$ 19.8_{-3.4}^{+3.2}$	\\[0.5ex]
   4 & 0.412 & 0.472 &$ 26.33_{-0.60}^{+0.80} $&$ 1.65_{-0.15}^{+0.12} $&$
	2.15_{-0.14}^{+0.60} $&$ 10.2_{-5.1}^{+4.8}$	\\[0.5ex]
   5 & 0.472 & 0.525 &$ 30.45_{-0.69}^{+0.96} $&$ 1.66_{-0.15}^{+0.13} $&$
	2.27_{-0.18}^{+0.76} $&$  6.8_{-5.3}^{+5.0}$	\\[0.5ex]
   6 & 0.525 & 0.582 &$ 33.02_{-0.75}^{+1.08} $&$ 1.51_{-0.12}^{+0.12} $&$
	2.18_{-0.19}^{+0.63} $&$  1.9_{-6.0}^{+6.0}$ 	\\[0.5ex]
   7 & 0.582 & 0.641 &$ 37.73_{-0.94}^{+1.37} $&$ 1.60_{-0.18}^{+0.07} $&$
	1.77_{-0.04}^{+0.47} $&$ 10.2_{-6.0}^{+5.6}$	\\[0.5ex]
   8 & 0.641 & 0.719 &$ 38.23_{-0.85}^{+1.19} $&$ 1.59_{-0.15}^{+0.08} $&$
	1.83_{-0.06}^{+0.38} $&$  7.6_{-5.5}^{+5.2}$	\\[0.5ex]
   9 & 0.719 & 1.543 &$ 37.28_{-0.93}^{+1.40} $&$ 1.43_{-0.12}^{+0.12} $&$
	2.04_{-0.16}^{+0.60} $&$ 13.1_{-7.5}^{+6.7}$	\\[2ex]
   - & 0.620 & 1.543 &$ 37.97_{-0.64}^{+0.81} $&$ 1.52_{-0.09}^{+0.08} $&$
	1.91_{-0.09}^{+0.24} $&$  9.8_{-4.1}^{+3.9}$	\\
 \hline
 \end{tabular}

 $\sigma_1$, $\sigma_2$, $\sigma_3$ are the roots of the largest, middle, and
 smallest eigenvalue of the velocity dispersion tensor $\bsq$. $l_v$ is
 the vertex deviation (Equation~\ref{vertex-dev}). Units are mag, \kms, and
 degree for $B$-$V$, $\sigma_i$, and $l_v$ respectively. The errors given
 correspond to the 15.7 and 84.3 percentiles, i.e.\ 1$\sigma$ error.
\end{table}

\subsection{The velocity dispersion tensor}
 Once the Sun's velocity $\bv_\odot$ with respect to the LSR is known, we can
determine proper motions $\bp'$ with respect to the LSR by replacing
$\ave{\bv}$ by $-\bv_\odot$ in Equation~(\ref{defspprime}). Then for any
stellar group we can determine $\bsq$ from these values of $\bp'$ as
described in Section~\ref{sec-tec-sig}.  We divided the 11\,865 stars of our
sample in nine bins in $B$-$V$ with equal numbers of stars in each bin. The
results are displayed in Figure~\ref{fig-sigma} and Table~\ref{tab-sig}.
The errors correspond to the 15.7 and 84.3 percentiles ($1\sigma$ error)
and have been evaluated assuming a multivariate Gaussian distribution in the
$\sigma_{ij}^2$ with variance evaluated via Equation~(\ref{sig-var}).
In the upper panel of Figure~\ref{fig-sigma} the  three diagonal velocity
dispersions and $\avet{v_\phi}$ are plotted versus $B$-$V$. Parenago's
discontinuity is visible in all three $\sigma_{ii}$ and $\avet{v_\phi}$,
though, due to the larger bin size, less clearly than in
Figure~\ref{fig-mean-motion} above. The ordering between the diagonal
components of $\bsq$ is the same for all colour bins: $\sigma_{xx} >
\sigma_{yy} > \sigma_{zz}$. 

 The lower three panels of Figure~\ref{fig-sigma} show
 \beq
 \sigma^\pr_{ij}\equiv\rmn{sign}(\sigma^2_{ij})\,\sqrt{|\sigma_{ij}|}
\eeq
 for the mixed components of the velocity dispersion tensor $\bsq$.
Evidently, the mixed moments involving vertical motions vanish within their
errors. This is to be expected for essentially all possible dynamical
configurations of the Milky Way. On the other hand, the mixed dispersion in
the plane, $\sigma^2_{xy}$ differs significantly from zero, which is not
allowed in a well-mixed axisymmetric Milky Way. Thus, the principal axes of
$\bsq$ are not aligned with our Cartesian coordinate frame. We diagonalized
the tensor $\bsq$ to obtain its eigenvalues $\sigma_i^2$. The square root of
the largest of these as well as the ratios to the smaller ones are given in
Table~\ref{tab-sig}.  The ratio $\sigma_1/\sigma_2\simeq1.6$, whereas
$\sigma_1/\sigma_3\simeq2.2$ with a trend for smaller values at redder
colours. 

Also given in Table~\ref{tab-sig} is the ``vertex deviation'', commonly used
to parameterize the deviation from dynamical symmetry. This is defined to be
 \beq \label{vertex-dev}
	\ell_\rmn v \equiv \fracj12 \arctan \biggl(
	{2\sigma^2_{xy}\over\sigma^2_{xx}-\sigma^2_{yy}}\biggr),
\eeq
 and is the angle by which one has to rotate our Cartesian coordinate system
around its $\hat{z}$ axis such that the resulting velocity dispersion tensor
is diagonal in the $v_xv_y$ plane. Hence $\ell_\rmn v$ is the Galactic
longitude of the direction of $\sigma_1^2$, the largest eigenvalue of the
velocity dispersion tensor.  Fig.~\ref{fig-vertex-dev} shows $\ell_\rmn v$
as a function of colour. Blueward of Parenago's discontinuity there is a
clear trend of $\ell_\rmn v$ decreasing with $B$-$V$. Redward of Parenago's
discontinuity stellar kinematics is independ of $B$-$V$ and it makes sense
to group all 3093 stars into a single bin. The results are given in the last
row of Table~\ref{tab-sig}. The hypothesis that $\ell_\rmn v\le0$ for this
group of stars is excluded at the 99 per cent confidence level.\footnote{
	It is important to bear in mind that the errors in $\ell_\rmn v$ are
	not distributed Gaussianly}

\ifREFEREE
\else
\begin{figure}
        \epsfxsize=21pc \epsfbox[18 480 620 720]{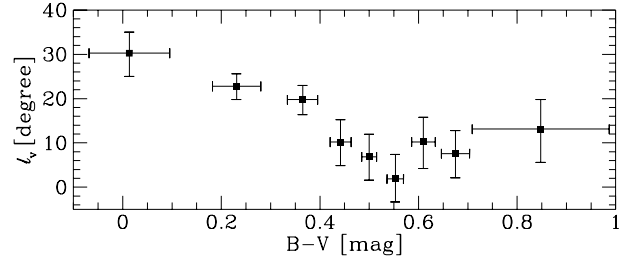}
 \caption[]{The vertex deviation $\ell_v$ versus $B$-$V$ colour. The error bars
	    correspond to the 15.7 and 84.3 percentile (i.e.\ 1$\sigma$ error)
	    and have been obtained assuming a multivariate Gaussian
	    distribution in the $\sigma_{ij}^2$ with variance evaluated
	    via Equation~(\ref{sig-var}).} \label{fig-vertex-dev}
\end{figure}
\fi

\section{DISUSSION}
We have analyzed a kinematically unbiased sample of nearly $12\,000$
main-sequence stars for which the \hip\ satellite measured parallaxes that
have errors smaller than 10 per cent. From this sample we have redetermined
most of the key kinematic parameters of the solar neighbourhood with 
unprecedented accuracy.

On account of the unusually small accidental and  systematic errors of the
\hip\ data, our errors are dominated by Poisson noise; we are characterizing
the distribution of stars in three-dimensional velocity space from samples of
typically 500 stars. From this fact it follows that our error estimates may
be considered reliable.

Equation (\ref{solar-vel}) gives the velocity of the Sun with respect to the
LSR. In common with several recent studies 
\cite{mayor74,gommes77,oblak83,biesec97} this has a smaller azimuthal
component ($V_0=5.2\pm0.6\kms$) than Delhaye's classical value, $12\kms$. A
contributory factor to this discrepancy is probably the fact that stars
bluer than $B$-$V\simeq0$ have anomalous values of both $\ave{v_x}$ and
$\ave{v_y}$, presumably because they are not yet well mixed. We have ignored
such stars in our determination of the Sun's motion with respect to the LSR.

Equation (\ref{Strom_reln}) quantifies the asymmetric drift. As is well
known, the coefficient $k$ that occurs in it constrains the radial gradients
of stellar density and the velocity-dispersion tensor, $\bsq$.
Quantitatively, if $R_\d$ is the scale-length of the disc, and $\nu$ the
mid-plane stellar density, with Equation (4-34) of BT we may infer from
Equation (\ref{Strom_reln}) that
 \ben\label{Gives_Rd}
{R_0\over R_\d} &\simeq& -\fracj12{\p\ln\nu\sigma_{xx}^2\over\p\ln R}
	\bigg|_{R_0} \nonumber\\
                &  =   & {v_c\over k}+\fracj12\bigg[1-
			 {\sigma_{yy}^2\over\sigma_{xx}^2} \pdot \Big(1-
			 {\sigma_{zz}^2\over\sigma_{xx}^2}\Big)\bigg]	\\
		& \simeq & 3.0\pdot0.4.\nonumber
\een
 Here the first equality involves the assumption that $\sigma_{xx}^2\propto\nu$,
which is based on the observation that disc scale heights vary little with 
radius, and the term that follows the symbol ${\pdot}$ may or may not be 
required, depending on whether the longest axis of $\bsq$ points to the 
galactic centre or remains horizontal as one moves above the plane [cf.\ 
\S4.2.1(a) of BT]. The numerical values given are based on Table~1, $R_0=8\kpc$
and the Feast \& Whitelock values of the Oort constants. The relatively short 
scale-length implied by Equation~(\ref{Gives_Rd}) is consistent with recent 
inrared-based studies of the Galaxy \cite{Kent,Spergeletal}. 

The shape of the velocity ellipsoid does not vary significantly with $B$-$V$;
to within the errors its axis ratios are constant at $\sigma_1/\sigma_3\simeq
2.2$ and $\sigma_1/\sigma_2\simeq1.6$. The ratio $\sigma_{yy}^2/\sigma_{xx}^2
\simeq0.4$ is intimately connected with the values of the Oort constants, 
being na\"\i vely given by Oort's equation $-B/(A-B)\simeq0.45$. Unfortunately,
Oort's relation is significantly in error for typical star samples 
\cite{Binney86} but Cuddeford \& Binney (1994) argue that $-B/(A-B)$ can be
accurately determined from $\sigma_{yy}^2/\sigma_{xx}^2$ if modified rather
than normal moments are employed. We hope soon to evaluate these modified 
moments from the present sample.

The ratio $\sigma_1/\sigma_3$ constrains the nature of the scattering
process that is responsible for the increase in stellar velocity dispersion
with age (e.g.~Jenkins 1992). The value determined here is consistent with
previous values, and implies that both spiral structure and scattering by
molecular clouds contribute significantly to heating of the local disc.  We
find a hint of smaller values of $\sigma_1/\sigma_3$ for redder colours,
which is the trend expected if heating by spiral structure is less efficient
relative to cloud-scattering for dynamically hotter stellar populations.

We have shown that one principal axis of the velocity ellipsoid coincides
with the direction $b=90\deg$. The angle by which the longest axis of the
velocity ellipsoid deviates from the direction to the Galactic centre, the
vertex deviation $\ell_\rmn v$, decreases from large values ($\ell_\rmn
v\gta20\deg$) for early types up to the colour of Parenago's discontnuity
and is then consistent with being constant at $\ell_\rmn v\simeq10\deg$. 

Two factors probably contribute to the vertex deviation: (i) the likelihood
that a significant fraction of young stars belong to a small number of
moving groups, and (ii) any large-scale non-axisymmetric component in the
Galactic potential, such as would be generated by either a bar or spiral
structure. Spiral arms will contribute to the vertex deviations of all
stellar groups, but their contribution will be largest for the populations
with the smallest velocity dispersions because stars with epicycle amplitudes
comparable to or larger than the interarm spacing will respond weakly to the
spiral's potential. Hence the more tightly wound the arms are, the more
strongly the effect will be confined to the earlier spectral types. The fact
that we see a significant vertex deviation to the latest spectral types,
implies that the causative agent is either a bar or rather open spiral arms.
To quantify this point, let $X$ be the epicycle amplitude of a typical
late-type star. Then the star's radius is $R(t)=R_\rmn g+X\cos(\kappa t)$,
where $\kappa=2\sqrt{B^2-AB}\simeq36.7\kms\kpc^{-1}$ is the epicycle frequency.
To a reasonable approximation we may equate the time average of ${\dot R}^2$ 
to $\sigma_1^2\simeq(38\kms)^2$. Hence $X\simeq\sqrt{2}\sigma_1/\kappa\simeq
1.5\kpc$ and the star moves $\sim3\kpc$ in each epicycle period. We would not
expect spiral arms with inter-arm spacing $\Delta\lta3\kpc$ to have much
effect on the orbit of such a  star. It is clearly important to quantify this
interesting conclusion more precisely.

Finally it seems worth remarking that in characterizing the distribution of
stars in velocity space by its first few moments we in no way imply that the
distribution is similar to Schwarzschild's ellipsoidal distribution, as
given, for example, by equation (7-91) of BT; the velocity ellipsoid is a
formal construct which need have no physical counterpart. From these data it
{\em is\/} possible to map the distribution of stars in velocity space and
discover how closely it resembles Schwarzschild's paradigm. But that topic
is reserved for a future paper (Dehnen, in preparation).

\end{document}